\documentstyle[epsfig]{aipproc}

\begin{document}
\title{Inflationary Reheating and Fermions}

\author{Patrick B. Greene}
\address{Department of Physics, University of Toronto,\\
60 St. George Street, Toronto, Ontario M5S 1A7, Canada
}

\maketitle

\begin{abstract}
Coherent oscillations of the inflaton field at the end of inflation
can parametrically excite fermions in much the same way that bosons
are created in preheating.  Although Pauli-blocking prohibits the 
occupation number of created fermions from growing exponentially, fermion
production  occurs in a manner significantly different from the
expectations of simple perturbation theory. Here,
I discuss the nature of fermion production after inflation and
possible applications including the efficient transfer of inflaton
energy and the production of super-massive fermions during
fermionic preheating.
\end{abstract}

	Consider a simple model of chaotic inflation with the potential 
${1 \over 4} \lambda \phi^4$  for an inflaton field $\phi$ coupled to
a massless spin-$1 \over 2$ field $\psi$ by a Yukawa interaction.
At the end of inflation, the inflaton
field will oscillate coherently about the minimum of its effective potential
with an initial amplitude $\phi_o \sim O(0.1 M_p)$.  
In perturbation theory, one treats the homogeneous and quasi-classical inflaton
field as a condensate of scalar inflaton particles, each of which can decay
into a pair of $\psi$-particles.  Each fermion then carries 
away half of the energy of a typical inflaton particle, giving a spectrum 
narrowly peaked around the comoving momentum $k \approx 0.42 \sqrt{\lambda} \phi_o$.
The inflaton energy is transfered to fermions 
after $O({1 \over h^2}) \gg 1$ inflaton oscillations.

	To investigate fermion production non-perturbatively, we are interested in the
equation of motion for the field operator $\psi$.  Following the usual prescription
(see \cite{GK98} and references therein for details), one seeks eigenfunctions
of the Dirac equation in the presence of the classical time-dependent source, $\phi(t)$.
As the inflaton field is spatially homogeneous, only the temporal part of the 
eigenmode obeys a non-trivial equation of motion. These modes, $X_k(\tau)$, obey 
an oscillator-type equation with a {\it complex} frequency that varies periodically
with time: 
\begin{equation}
X_k''  +  {\left(\kappa^2  +
 q f^2 -i   \sqrt{q} f' 
 \right)} X_k  = 0 \ .
\label{mode}
\end{equation}
Here, the comoving momentum $k$ enters the equation in the combination
 ${\kappa^2 \equiv {k^2 \over  {\lambda \phi_o^2}}}$ and the character of the 
solutions is defined by the parameter $q \equiv {h^2 \over \lambda}$ ($h$ is
the dimensionless Yukawa coupling).
The background oscillations enter in the form
 $f(\tau)=cn \left( \tau, { 1 \over \sqrt{2}}\right)$ having
unit amplitude and a period $T=7.416$.  Note that we are working in scaled conformal
field and time variables so the effects of expansion do not appear in the equations
of motion.
We can express
the comoving occupation number of  particles in a given state
through the solutions of eq.~(\ref{mode}) 
\begin{equation}
n_k(\tau) = {{(\Omega_k - \sqrt{q}f)} \over {2\Omega_k}}[~|X_k'|^2 + \Omega_k^2 |X_k|^2
- 2\Omega_k Im(X_k {X_k'}^*)~], 
\label{nk}
\end{equation}
where $\Omega_k^2 \equiv \kappa^2 + q f^2$.
The energy density of created fermions is
$\epsilon_{\psi}={ 1\over 2\pi^3 }\int d^3k~ \Omega_k~ n_k$.
\begin{figure}[tb]
\begin{minipage}[t]{7.2cm}
   \centering \leavevmode \epsfxsize=7.2cm
   \epsfbox{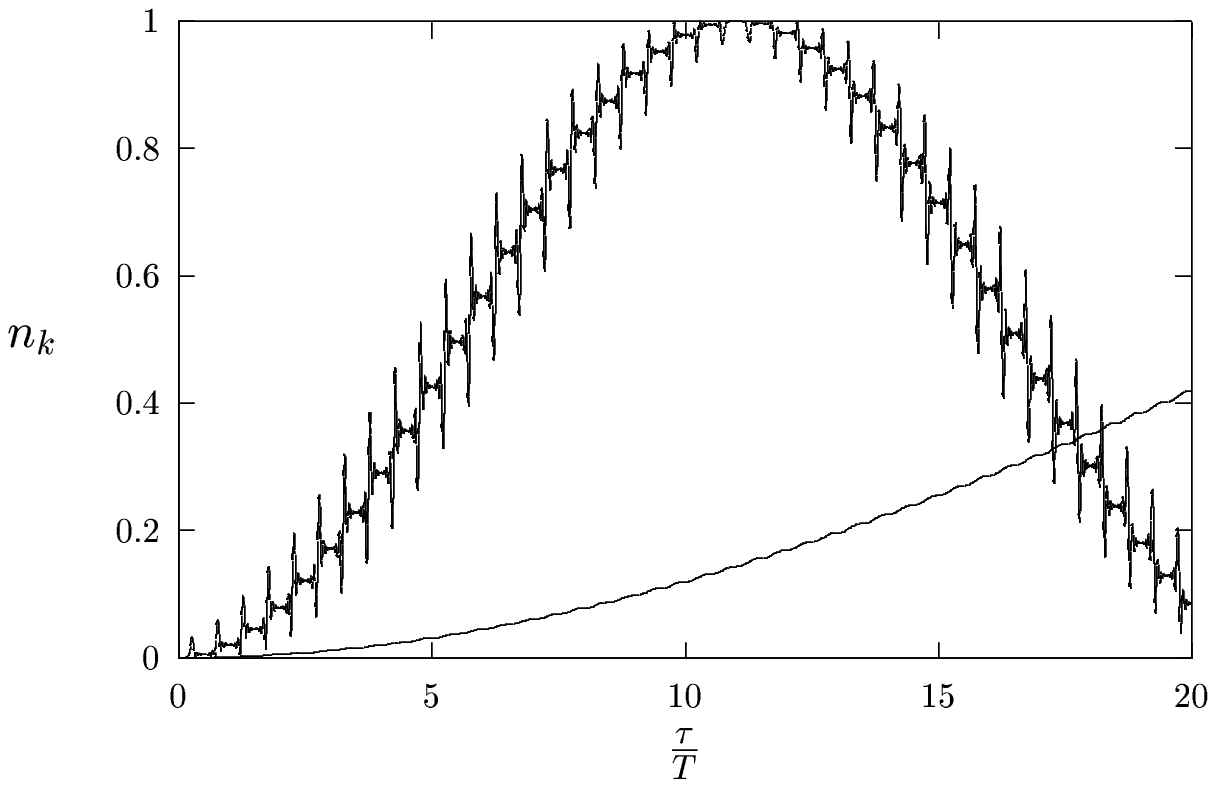}\\

   \caption[fig1u]{\label{pbg:f:tdev} {\em 
The comoving occupation number $n_k$ of fermions in 
$\lambda \phi^4$-inflation as a function of time 
(in units of inflaton oscillations)  for 
$q \equiv {h^2 \over \lambda} = 10^{-4}$ and $100$ and 
$\kappa^2 = 0.18$ and $11.9$, respectively.
The period of the  modulation 
$\pi \over {\nu_k T}$ is about $88$ and  $22$ accordingly.
\hspace*{\fill}}}
\end{minipage}
\hfill
\begin{minipage}[t]{7.2cm}
   \centering \leavevmode \epsfxsize=7.2cm
   \epsfbox{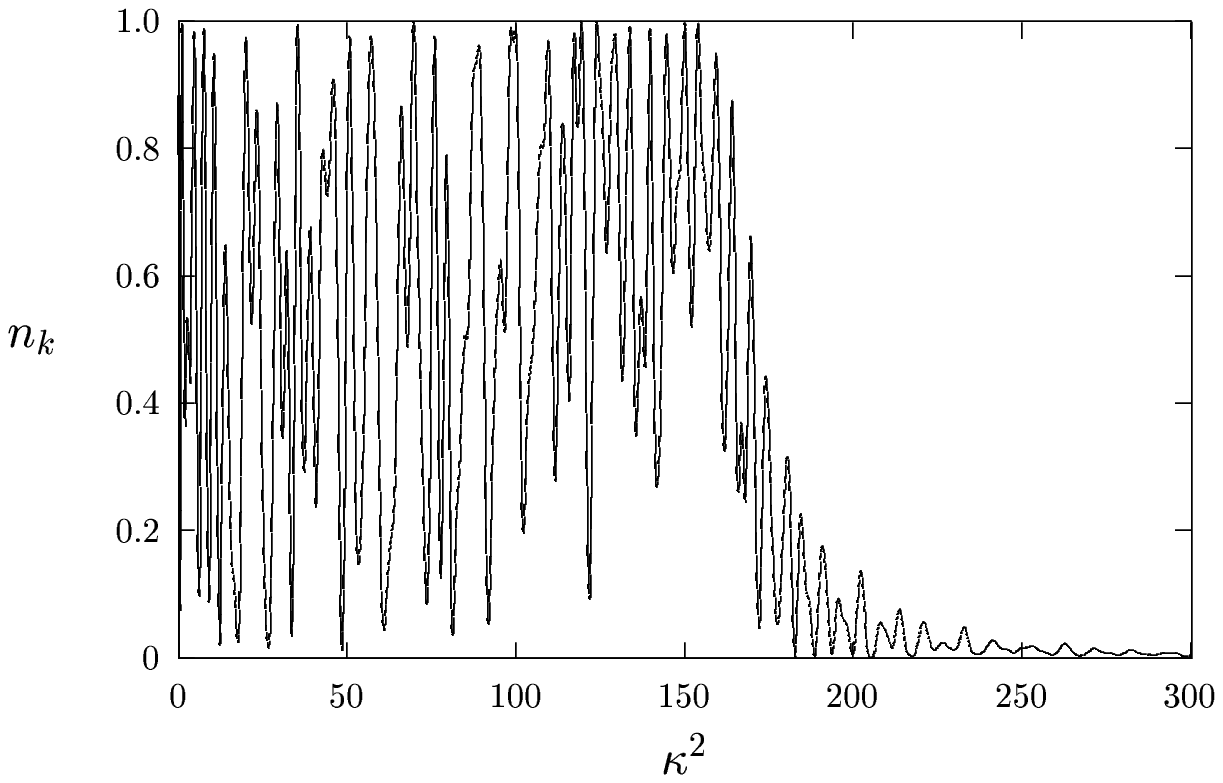}\\

   \caption[fig1u]{\label{pbg:f:sphere} {\em 
The comoving occupation number of fermions in $m_\phi^2 \phi^2$-inflation 
as a function of (scaled) comoving momentum after $50$ inflaton 
oscillations for initial resonance parameter $q_o = 10^3$.
Expansion destroys the details of the resonance band and
leads to a fermi-sphere of width $q^{1/4} \sim {q_o^{1/4}}a^{1/4}$.
  \hspace*{\fill}}}
\end{minipage}
\end{figure}

\begin{figure}[tb]
\begin{minipage}[t]{7.2cm}
   \centering \leavevmode \epsfxsize=7.2cm
   \epsfbox{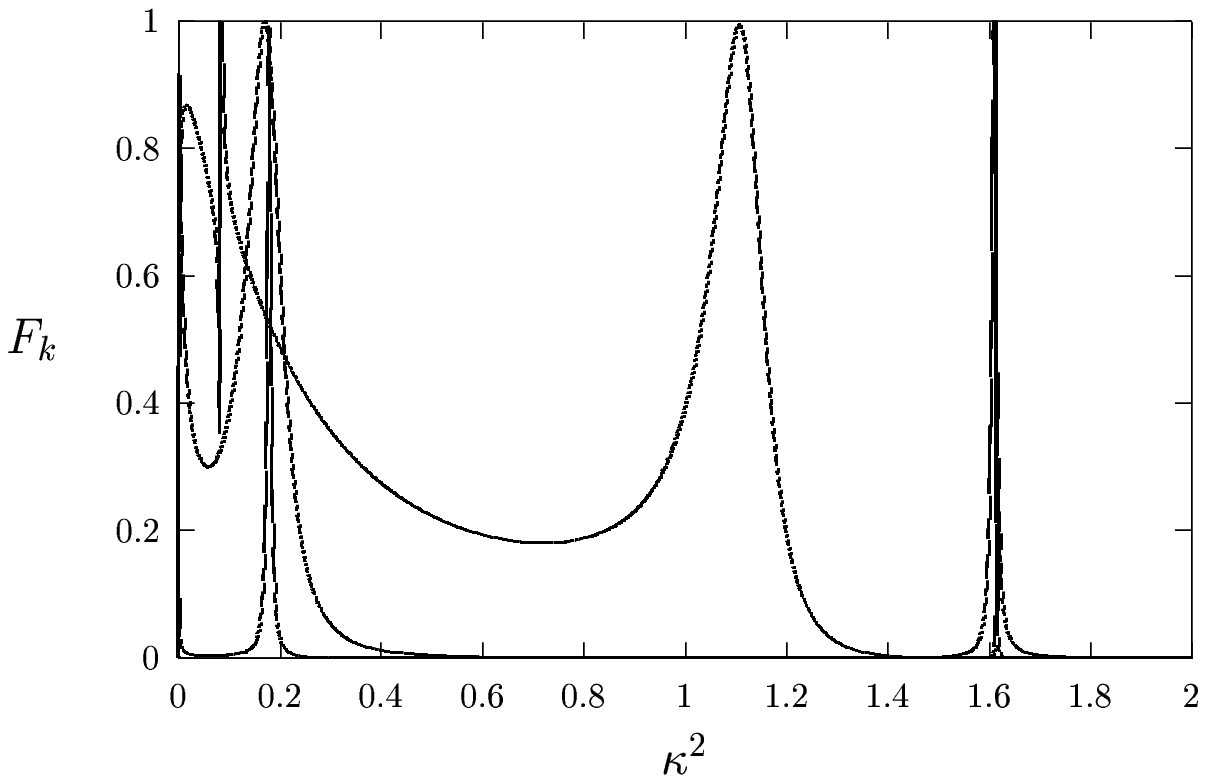}\\

   \caption[fig1u]{\label{pbg:f:envelope} {\em 
The envelope functions $F_k$ showing the bands of fermion
resonance excitation in $\lambda \phi^4$-inflation for   
$q \equiv {h^2 \over \lambda} = 10^{-4}, 10^{-2}$, and $1.0$
(the narrowest to broadest band, respectively).
\hspace*{\fill}}}
\end{minipage}
\hfill
\begin{minipage}[t]{7.2cm}
   \centering \leavevmode \epsfxsize=7.2cm
   \epsfbox{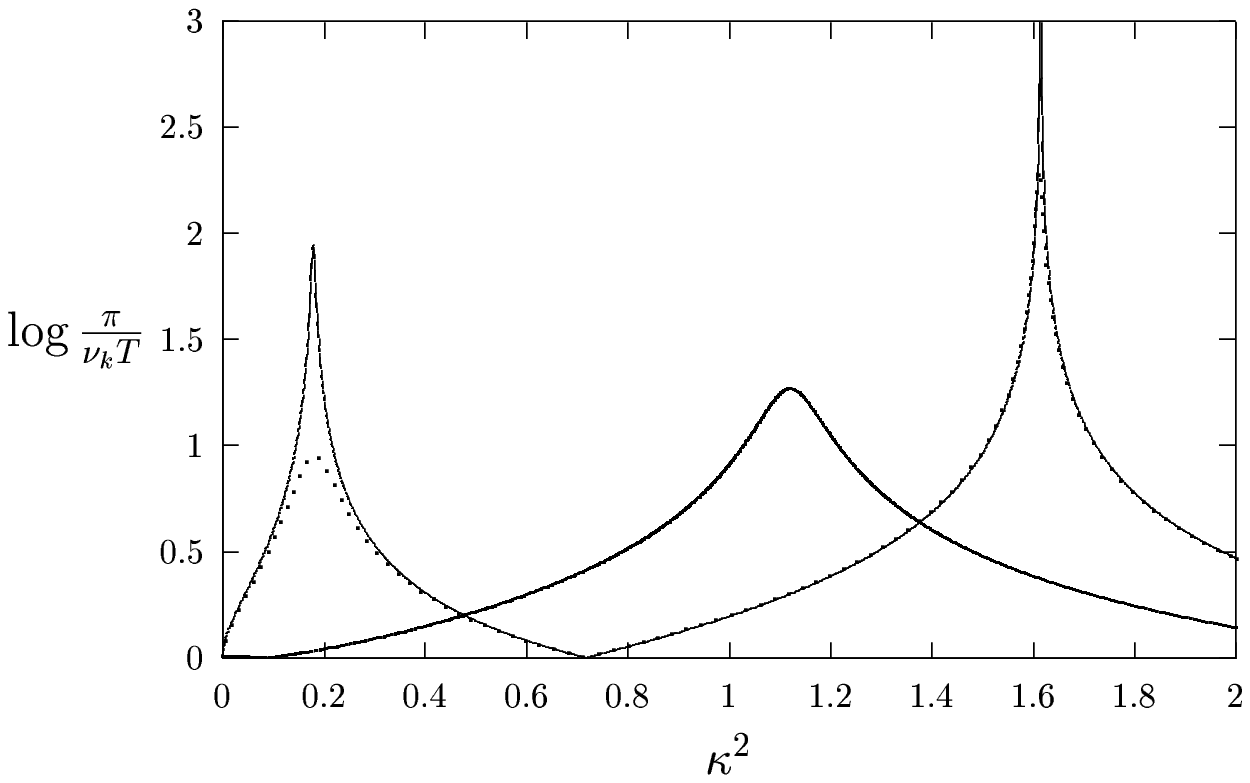}\\

   \caption[fig1u]{\label{pbg:f:period} {\em 
The $\log$ of the period of modulation ${\pi \over {\nu_k T}} $
(in units of inflaton oscillations) as a function of $\kappa^2$ for 
$q \equiv {h^2 \over \lambda} = 10^{-4}, 10^{-2}$, and $1.0$
for the light, dotted, and heavy  curves respectively.
  \hspace*{\fill}}}
\end{minipage}
\end{figure}

	It turns out that, for all $\kappa^2$ and $q$, the solutions of eq.~(\ref{mode})
are periodic in time.  This is shown by the numerical solutions for the comoving 
occupation number in Fig.~(\ref{pbg:f:tdev}).  Furthermore, it is easy to show
that the comoving occupation number defined by eq.~(\ref{nk}) will obey $n_k \le 1$
in accordance with the Pauli principle.  We see from Fig.~(\ref{pbg:f:tdev}) that,
while the occupation number exhibits some high frequency oscillations
(period $< {T \over 2}$), the most interesting behavior occurs over longer periods.
If we average the occupation number over an inflaton period,
$\bar n_k(\tau)= { 1 \over T} \int_{\tau}^{(\tau+T)} d \tau  n_k(\tau)$,
the average occupation number is found to obey the simple equation:
$\bar n_k(\tau) = F_k \sin^2  \nu_k \tau$.  For a given theory, i.e. for
a given value of the resonance parameter $q$, $F_k \le 1$ is a
momentum dependent amplitude and $\nu_k$ is a momentum dependent frequency.
In Fig.~(\ref{pbg:f:envelope}) the amplitude $F_k$ as a function of $\kappa^2$ is plotted
for several values of $q$.  We see that the perturbative expectation is 
only met for $q \le 10^{-4}$.

	In fact, for large $q$, the fermions are excited up to $\kappa^2 \simeq \sqrt{q}$.
This is the same result as for the broad bosonic resonance and can be understood in
the same manner.  The comoving occupation number, eq.~(\ref{nk}),
is an adiabatic invariant of the mode equation (\ref{mode}).  The condition that
the modes $X_k$ evolve non-adiabatically is  $\dot \Omega_k \geq \Omega_k^2$
which leads to the condition $\kappa^2 \le \sqrt{q}$ for non-adiabatic evolution, 
and thus, particle creation.  
This occurs much more rapidly than one would expect from
perturbation theory.  In Fig.~(\ref{pbg:f:period}) the period of mode oscillations
${\pi \over {\nu_k T}}$ is plotted in units of inflaton oscillations.  For
all $q$, the bands saturate after only $10-100$ inflaton oscillations.  

Turning to the energetics for the most interesting case of the broad resonance
excitation, $q \gg 1$, we find $\epsilon_{\psi} \sim 0.1 h^2 q^{1/4} \epsilon_{\phi}$,
where the inflaton energy is
$ \epsilon_{\phi}={1 \over 4} \lambda \phi_o^4$.
In chaotic ${1 \over 4} \lambda \phi^4$-inflation, $\lambda \simeq 10^{-13}$.
If the resonance parameter $q$ is large but the coupling parameter
is small, $h \leq 0.1$, only a small fraction of the inflaton's energy will be
converted into fermions.  Although explosive decay will not occur for this model, 
more general theories such as hybrid models can have large enough resonance 
parameters to allow efficient decay to fermions.

If the fermion field has a small bare mass term or if we consider inflation
with a ${1 \over 2} m_\phi^2 \phi^2$ potential, the conformal invariance of
the theory will be broken.  In this case, the occupation number of fermions
for $q \gg 1$ no longer evolves periodically but becomes stochastic, rapidly fluctuating 
between $n_k=1$ and $n_k=0$.  This destroys the well defined resonance bands
depicted in Fig.~(\ref{pbg:f:envelope}) and can lead to the production of
super-massive fermions of mass $m_\psi \le h \phi_o$.  These fermions can come
to dominate the energy density of the universe or survive as massive relics.

	As a particular example of the changes brought by expansion, 
consider $m_\phi^2  \phi^2$-inflation with a Yukawa coupling to a still
massless fermion.  In the broad resonance case, one gets a sphere in 
comoving momentum space with average occupation $\bar n_k ={1 \over 2}$ 
that {\it expands} with time.  A snapshot of this sphere is shown in 
Fig.~(\ref{pbg:f:sphere}).

\end{document}